%% file: paper.tex
\documentclass{article}

\usepackage{aaai}

\usepackage{times}
\usepackage{helvet}
\usepackage{courier}

\usepackage[pdftex]{graphicx}
\usepackage{amsmath, amsfonts, amsthm, amssymb}
\usepackage{xspace}

\newcommand{\Figure}[4]{
\begin{figure}[t]
\centering
\includegraphics[width=#2]{#1}
\caption{#4}
\label{fig:#3}
\end{figure}
}

\newcommand{\Mff}  {Metric-FF\xspace}
\newcommand{\Sg}   {SGPlan\xspace}

\title{Attack Planning in the Real World}
\author{Jorge Lucangeli Obes \\
Core Security Technologies \\
{\em jlucangelio@coresecurity.com}
\And Carlos Sarraute \and Gerardo Richarte \\
Core Security Technologies \\
and Instituto Tecnologico Buenos Aires \\
{\em \{carlos, gera\}@coresecurity.com}
}

\nocopyright

\begin{document}
\maketitle

\begin{abstract}
Assessing network security is a complex and difficult task. Attack graphs
have been proposed as a tool to help network administrators understand
the potential weaknesses of their networks. However, a problem has not yet
been addressed by previous work on this subject; namely, how to actually
execute and validate the attack paths resulting from the analysis of
the attack graph. In this paper we present a complete PDDL representation
of an attack model, and an implementation that integrates a planner into
a penetration testing tool. This allows to automatically generate
attack paths for penetration testing scenarios, and to validate these attacks
by executing the corresponding actions -including exploits- against the
real target network. We present an algorithm for transforming the information
present in the penetration testing tool to the planning domain, and we show
how the scalability issues of attack graphs can be solved using current
planners. We include an analysis of the performance of our solution, showing
how our model scales to medium-sized networks and the number of actions
available in current penetration testing tools.
\end{abstract}

\section{Introduction}
\label{sec:introduction}

The last 10 years have witnessed the development of a new kind of information
security tool: the penetration testing framework. These tools facilitate
the work of network penetration testers, and make the assessment of network
security more accessible to non-experts. The main tools available are the
open source project Metasploit, and the commercial products Immunity Canvas
and Core Impact \cite{secpowertools}.

The main difference between these tools and network security scanners such
as Nessus or Retina is that pentesting frameworks have the ability to launch
real exploits for vulnerabilities, helping to expose risk by conducting
an attack in the same way a real external attacker would \cite{arce-attacking}.

Penetration tests involve successive phases of \emph{information gathering},
where the pentesting tool helps the user gather information about the network
under attack (available hosts, their operating systems and open ports,
and the services running in them); and exploiting, where the user actively
tries to compromise specific hosts in the network. When an exploit launched
against a vulnerable machine is successful, the machine becomes compromised
and can be used to perform further information gathering, or to launch
subsequent attacks. This shift in the source of the attacker's actions
is called \emph{pivoting}.
Newly compromised machines can serve as the source for posterior information
gathering, and this new information might reveal previously unknown vulnerabilities,
so the phases of information gathering and exploiting usually succeed one
another. 

As pentesting tools have evolved and have become more complex, covering new
attack vectors; and shipping increasing numbers of exploits and information gathering modules,
the problem of controlling the pentesting framework successfully became an important question.
A computer-generated plan for an attack would isolate the user from the
complexity of selecting suitable exploits for the hosts in the target network.
In addition, a suitable model to represent these attacks would help to
systematize the knowledge gained during manual penetration tests performed
by expert users, making pentesting frameworks more accessible to non-experts.
Finally, the possibility of incorporating the attack planning phase to
the pentesting framework would allow for optimizations based on exploit
running time, reliability, or impact on Intrusion Detection Systems.

Our work on the attack planning problem applied to pentesting began in 2003
with the construction
of a conceptual model of an attack, distinguishing assets, actions and
goals \cite{building2003,gera2003,pacsec2003}. In this attack model,
the assets represent both information and the modifications in the network
that an attacker may need to obtain during an intrusion, whereas the actions
are the basic steps of an attack, such as running a particular exploit
against a target host. This model was designed to be realistic from an
attacker's point of view, and contemplates the fact that the attacker has
an initial incomplete knowledge of the network, and therefore information
gathering should be considered as part of the attack.

Since the actions have requirements (preconditions) and results, given a
goal, a graph of the actions/assets that lead to this goal can be
constructed. This graph is related to the {\em attack graphs}\footnote{Nodes
in an attack graph identify a stage of the attack, while edges represent
individual steps in the attack.} studied in \cite{phillips1998,jajodia2005,noel2009}
and many others. In \cite{lippmann2005} the authors reviewed past papers
on attack graphs, and observed that the ``first major limitation of these
studies is that most attack graph algorithms have only been able to generate
attack graphs on small networks with fewer than 20 hosts''.

In medium-sized networks, building complete attack graphs quickly becomes
unfeasible (their size increases exponentially with the number of machines
and available actions). To deal with the attack planning problem, 
a proposed approach \cite{pacsec2008,phrack2009} is to translate the model
into a PDDL representation and use classical planning algorithms to find attack paths.
Planning algorithms manage to find paths in the attack graph without constructing
it completely, thus helping to avoid the combinatorial explosion \cite{BlumFurst97}.
A similar approach was presented at SecArt'09 \cite{ghosh09}, but the authors'
model is less expressive than the one used in this work, as their objective
was to use the attack paths to build a minimal attack graph, and not to
carry out these attacks against real networks.

In this paper we present an implementation of these ideas. We have developed
modules that integrate a pentesting framework with an external planner,
and execute the resulting plans back in the pentesting framework, against
a real network. We believe our implementation proves the feasability of
automating the attack phases of a penetration test, and allows to think
about moving up to automate the whole process. We show how our model, and
its PDDL representation, scale to hundreds of target nodes and available
exploits, numbers that can be found when assessing medium-sized networks
with current pentesting frameworks.

The paper is structured as follows: in Section \ref{sec:architecture} we
present a high-level description of our solution, describing the steps needed
to integrate a planner with a penetration testing framework. Section \ref{sec:pddl}
describes our PDDL representation in detail, explaining how the ``real world''
view that we take forces a particular implementation of the attack planning
problem in PDDL. Section \ref{sec:testing} presents the results of our
scalability testing, showing how our model manages medium-sized networks
using current planners. Section \ref{sec:related} reviews related work,
and Section \ref{sec:conclusion} closes the paper and outlines future work.

\section{Architecture of our Solution}
\label{sec:architecture}

In this section we describe the components of our solution, and how
they fit together to automate an attack. Figure \ref{fig:architecture}
shows the relationship between these different components. The {\em penetration
testing framework} is a tool that allows the user/attacker to execute exploits
and other pre/post exploitation modules against the target network. Our
implementation is based on Core Impact\footnote{As mentioned in the previous
section, Metasploit is an open-source alternative.}. The \emph{planner}
is a tool that takes as input the description of a \emph{domain} and a
\emph{scenario}, in PDDL\footnote{Refer to \cite{pddl2} for a description
of the PDDL planning language.}. The domain contains the definition of
the available actions in the model, and the scenario contains the definition
of the objects (networks, hosts, and their characteristics), and the goal
which has to be solved.

\Figure{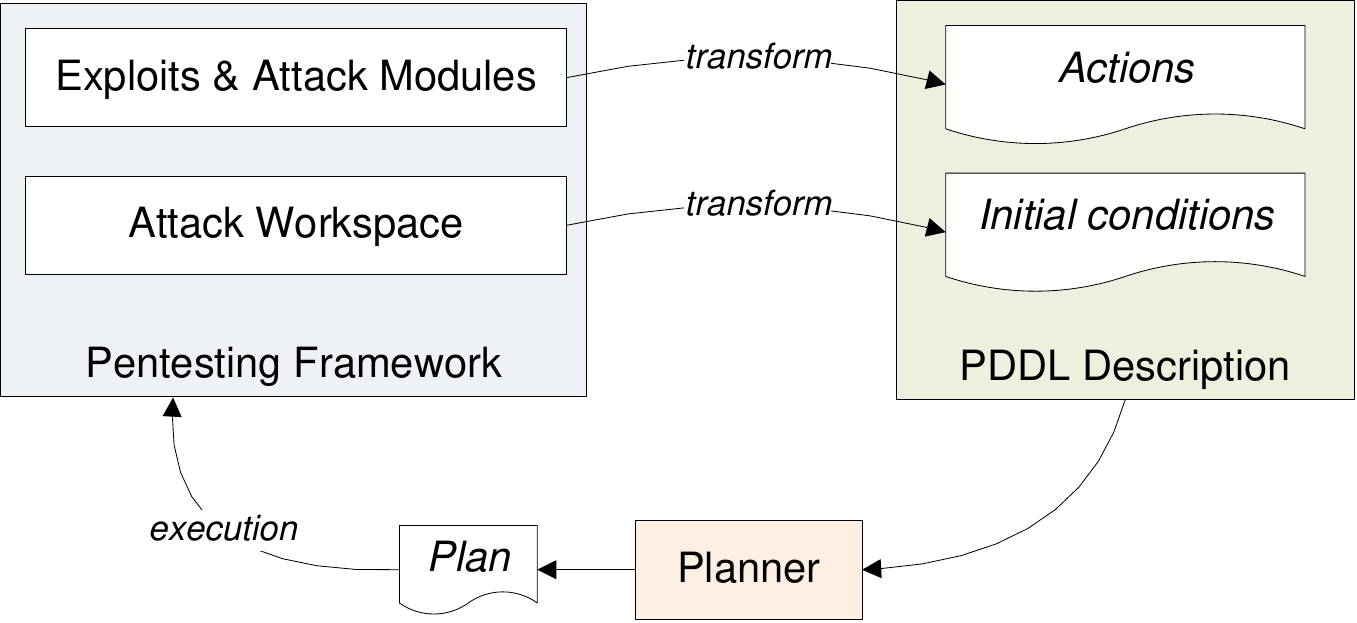}{8cm}{architecture}{Architecture of our solution.}

The {\em attack workspace} contains the information about the current attack
or penetration test. In particular, the discovered networks and hosts,
information about their operating systems, open/closed ports, running services
and compromised machines. In the current version of our solution we assume
that the workspace has this network information available, and that no
network information gathering is needed to generate a solvable plan. We
will address this limitation in Section \ref{sec:conclusion} when we discuss
future work.

\subsection{Transform algorithm}

The {\em transform} algorithm generates the PDDL representation of the
attack planning problem, including the initial conditions, the operators
(PDDL actions), and the goal. From the pentesting framework we extract the description of
the operators, in particular the requirements and results of the exploits,
which will make up most of the available actions in the PDDL representation.
This is encoded in the {\em domain.pddl} file, along with the predicates
and types (which only depend on the details of our model).

From the attack workspace we extract the information that constitutes the
initial conditions for the planner: networks, machines, operating systems,
ports and running services. This is encoded in the {\em problem.pddl} file,
together with the goal of the attack, which will usually be the \emph{compromise}
of a particular machine.

A common characteristic of pentesting frameworks is that they
provide an incomplete view of the network under attack. The pentester has
to infer the structure of the network using the information that he sees
from each compromised machine. The \emph{transform} algorithm takes this
into account, receiving extra information regarding host connectivity.

\subsection{Planner}

The PDDL description is given as input to the {\em planner}.
The advantage of using the PDDL language is that we can experiment with
different planners and determine which best fits our particular problem.
We have evaluated our model using both \Sg \cite{sgplan} and \Mff \cite{metric-ff}.

The planner is run from inside the pentesting framework, as a pluggable
module of the framework that we call \emph{PlannerRunner}. The output of
the planner is a {\em plan}, a sequence of actions that lead
to the completion of the goal, if all the actions are successful. We make
this distinction because even with well-tested exploit code, not all exploits
launched are successful. The plan is given as feedback to the pentesting
framework, and executed against the real target network.

\section{The PDDL Representation in Detail}
\label{sec:pddl}

The PDDL description language serves as the bridge between the pentesting
tool and the planner. Since exploits have strict platform and connectivity
requirements, failing to accurately express those requirements in the PDDL
model would result in plans that cannot be executed against real networks.
This forces our PDDL representation of the attack planning problem to be
quite verbose.

On top of that, we take advantage of the optimization abilities of planners
that understand numerical effects\footnote{Numerical effects allow the
actions in the PDDL representation to increase the value of different metrics
defined in the PDDL scenario. The planner can then be told to find a plan
that minimizes a linear function of these metrics.}, and have the PDDL
actions affect different metrics commonly associated with penetration
testing such as running time, probability of success or possibility of
detection (stealth).

We will focus on the description of the {\em domain.pddl} file, which
contains the PDDL representation of the attack model. We will not delve
into the details of the {\em problem.pddl} file, since it consists of a
list of networks and machines, described using the predicates to be presented
in this section.

The PDDL requirements of the representation are {\bf :typing}, so that
predicates can have types, and {\bf :fluents}, to use numerical effects.
We will first describe the types available in the model, and then list
the predicates that use these types. We will continue by describing the
model-related actions that make this predicates true, and then we will
show an example of an action representing an exploit. We close this section
with an example PDDL plan for a simple scenario.

\subsection{Types}

Table \ref{tab:types} shows a list of the types that we used. Half of
the object types are dedicated to describing in detail the operating
systems of the hosts, since the successful execution of an exploit
depends on being able to detect the specifics of the OS.

\begin{table}[ht]
\begin{center}
\scriptsize
\begin{tabular}{|l|l|}
\hline
network     & operating\_system \\
host        & OS\_version \\
port        & OS\_edition \\
port\_set   & OS\_build \\
application & OS\_servicepack \\
agent       & OS\_distro \\
privileges  & kernel\_version \\
\hline
\end{tabular}
\normalsize
\end{center}
\caption{List of object types}
\label{tab:types}
\end{table}

\subsection{Predicates}

The following are the predicates used in our model of attacks. Since exploits also
have non-trivial connectivity requirements, we chose to have a detailed
representation of network connectivity in PDDL. We need to be able to express
how hosts are connected to networks, and the fact that exploits may need
both IP and TCP or UDP connectivity between the source and target hosts, usually on
a particular TCP or UDP port. These predicates express the different forms
of connectivity:

{\scriptsize 
\begin{verbatim} 
(connected_to_network ?s - host ?n - network)
(IP_connectivity ?s - host ?t - host)
(TCP_connectivity ?s - host ?t - host ?p - port)
(TCP_listen_port ?h - host ?p - port)
(UDP_listen_port ?h - host ?p - port)
\end{verbatim} }

These predicates describe the operating system and services of a host:

{\scriptsize 
\begin{verbatim}
(has_OS ?h - host ?os - operating_system)
(has_OS_version ?h - host ?osv - OS_version)
(has_OS_edition ?h - host ?ose - OS_edition)
(has_OS_build ?h - host ?osb - OS_build)
(has_OS_servicepack ?h - host ?ossp - OS_servicepack)
(has_OS_distro ?h - host ?osd - OS_distro)
(has_kernel_version ?h - host ?kv - kernel_version)
(has_architecture ?h - host ?a - OS_architecture)
(has_application ?h - host ?p - application)
\end{verbatim} }

\subsection{Actions}

We require some ``model-related'' actions that make true the aforementioned
predicates in the right cases.

{\scriptsize 
\begin{verbatim}
(:action IP_connect
:parameters (?s - host ?t - host)
:precondition (and (compromised ?s)
  (exists (?n - network)
    (and (connected_to_network ?s ?n) 
      (connected_to_network ?t ?n))))
:effect (IP_connectivity ?s ?t)
)

(:action TCP_connect
:parameters (?s - host ?t - host ?p - port)
:precondition (and (compromised ?s)
  (IP_connectivity ?s ?t)
  (TCP_listen_port ?t ?p))
:effect (TCP_connectivity ?s ?t ?p)
)

(:action Mark_as_compromised
:parameters (?a - agent ?h - host)
:precondition (installed ?a ?h)
:effect (compromised ?h)
)
\end{verbatim}
}

Two hosts on the same network possess IP connectivity, and two hosts have
TCP (or UDP) connectivity if they have IP connectivity and the target host
has the correct TCP (or UDP) port open. Moreover, when an exploit is successful an
\emph{agent} is installed on the target machine, which allows control over
that machine. An installed agent is hard evidence that the machine is vulnerable,
so it marks the machine as compromised\footnote{Depending on the exploit
used, the agent might have regular user privileges, or superuser ({\em root})
privileges. Certain local exploits allow a low-level (user) agent to be
upgraded to a high-level agent, so we model this by having two different
\emph{privileges} PDDL objects.}.

The penetration testing framework we used has an extensive test suite that collects
information regarding running time for many exploit modules. We obtained
average running times from this data and used that information as the numeric
effect of exploit actions in PDDL. The metric to minimize in our PDDL scenarios
is therefore the total running time of the complete attack.

Finally, this is an example of an action, an exploit that will attempt
to install an agent on target host {\em t} from an agent previously installed
on the source host {\em s}. To be successful, this exploit requires that
the target runs a specific OS, has the service {\em ovtrcd} running and
listening on port 5053.

{\scriptsize 
\begin{verbatim}
(:action HP_OpenView_Remote_Buffer_Overflow_Exploit
:parameters (?s - host ?t - host)
:precondition (and (compromised ?s)
  (and (has_OS ?t Windows)
    (has_OS_edition ?t Professional)
    (has_OS_servicepack ?t Sp2)
    (has_OS_version ?t WinXp)
    (has_architecture ?t I386))
  (has_service ?t ovtrcd)
  (TCP_connectivity ?s ?t port5053)
)
:effect(and (installed_agent ?t high_privileges)
 	(increase (time) 10)
))
\end{verbatim} }

In our PDDL representation there are several versions of this exploit, one
for each specific operating system supported by the exploit. For example,
another supported system for this exploit looks like this:

{\scriptsize 
\begin{verbatim}
(and (has_OS ?t Solaris)
  (has_OS_version ?t V_10)
  (has_architecture ?t Sun4U))
\end{verbatim} }

The main part of the {\em domain.pddl} file is devoted to the description
of the actions. In our sample scenarios, this file has up to 28,000
lines and includes up to 1,800 actions. The last part of the {\em domain.pddl}
file is the list of constants that appear in the scenario, including the
names of the applications, the list of port numbers and operating system
version details.

\subsection{An attack plan}

We end this section with an example plan obtained by running \Mff on a
scenario generated with this model. The goal of the scenario is to compromise
host 10.0.5.12 in the target network. This network is similar to the test
network that we will describe in detail in Section \ref{sec:testing}. The plan
requires four pivoting steps and executes five different exploits
in total, though we only show the first\footnote{The \emph{localagent} object
represents the pentesting framework running in the machine of the user/attacker.}
and last ones for space reasons. The exploits shown are real-world exploits
currently present in the pentesting framework.

\input{plan.tex}

\section{Performance and Scalability Evaluation}
\label{sec:testing}

This model, and its representation in PDDL, are intended to be used to
plan attacks against real networks, and execute them using a pentesting
framework. To verify that our proposed solution scales up to the domains
and scenarios we need to address in real-world cases, we carried out extensive
performance and scalability testing -- to see how far we could take the
attack model and PDDL representation with current planners. We focused
our performance evaluation on four metrics:

\begin{itemize}

\item Number of machines in the attacked network

\item Number of pivoting steps in the attack

\item Number of available exploits in the pentesting suite

\item Number of individual predicates that must be fulfilled to accomplish the goal

\end{itemize}

The rationale behind using these metrics is that we needed our solution
to scale up reasonably with regard to all of them. For example, a
promising use of planning algorithms for attack planning lies in
scenarios where there are a considerable number of machines to
take into account, which could be time-consuming for a human attacker.

Moreover, many times a successful penetration test needs to reach the
innermost levels of a network, sequentially exploiting many machines in
order to reach one which might hold sensitive information.
We need our planning solution to be able to handle these cases where
many pivoting steps are needed.

Pentesting suites are constantly updated with exploits for new
vulnerabilities, so that users can test their systems against the latest
risks. The pentesting tool that we used currently\footnote{As of March, 2010.}
has about 700 exploits, of which almost 300 are the remote exploits that
get included in the PDDL domain. Each remote exploit is represented as
a different operator for each target operating system, so our PDDL domains
usually have about 1800 operators, and our solution needs to cope with
that input.

Finally, another promising use of planning algorithms for attack planning
is the continuous monitoring of a network by means of a constant pentest.
In this case we need to be able to scale to goals that involve compromising
many machines.

We decided to use the planners \Mff\footnote{Latest version available (with additional improvements).} \cite{metric-ff} 
and \Sg\footnote{\Sg version 5.22.} \cite{sgplan} since we
consider them to be representative of the state of the art in classical planners.
The original FF planner was the baseline planner for IPC'08\footnote{The
International Planning Competition, 2008.}. \Mff adds numerical effects
to the FF planner. We modified the reachability analysis in \Mff
to use type information, as in FF, to obtain better memory usage.

\Sg combines \Mff as a base planner with a constraint partitioning scheme
which allows it to divide the main planning problem in subproblems; these
subproblems are solved with a modified version of \Mff, and the individual
solutions combined to obtain a plan for the original problem. This method,
according to the authors, has the potential to significantly reduce the
complexity of the original problem \cite{sgplan}. It was successfully
used in \cite{ghosh09}.

\subsection{Generating the test scenarios}

We tested both real and simulated networks, generating the test scenarios
using the same pentesting framework we would later use to attack them. For the
large-scale testing, we made use of a network simulator \cite{simutools09}.
This simulator allows to build sizable networks\footnote{We tested up to
1000 nodes in the simulator.}, but still view each machine independently
and, for example, execute distinct system calls in each of them. The simulator
integrates tightly with the pentesting framework, to the point where the
framework is oblivious to the fact that the network under attack is simulated
and not real.

This allowed us to use the pentesting tool to carry out all the steps of
the test, including the information gathering stage of the attack. Once
the information gathering was complete, we converted the attack workspace
to PDDL using our \emph{transform} tool.

\Figure{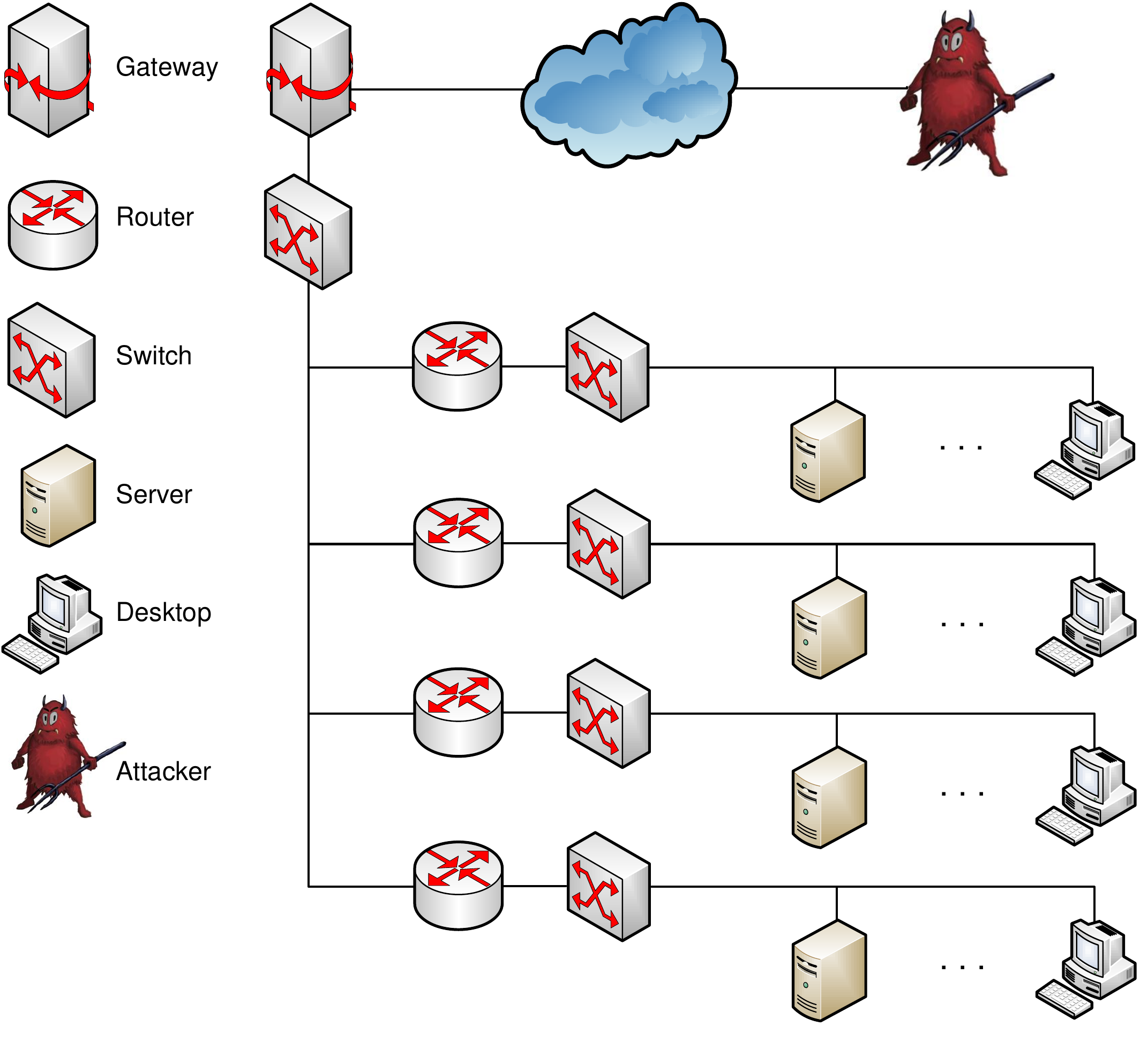}{8cm}{test_net}{Test network for scalability evaluation.}

We generated two types of networks for the performance evaluation. To
evaluate the scalability in terms of number of machines, number of
operators, and number of goals; the network consists of five subnets with
varying numbers of machines, all joined to one main network to which the
user/attacker initially has access. Figure \ref{fig:test_net} shows the
high-level structure of this simulated network.

To evaluate the scalability in terms of the number of pivoting steps
needed to reach the goal, we constructed a test network where the attacker
and the target machine are separated by an increasing number of routers,
and each subnetwork in between has a small number of machines.

\begin{table}[ht]
\begin{center}
\scriptsize
\begin{tabular}{|l|c|c|c|}
\hline
\textbf{Machine type} & \textbf{OS version} & \textbf{Share} & \textbf{Open ports} \\
\hline
Windows desktop & Windows XP SP3 & 50\% & 139, 445 \\
\hline
Windows server & Windows 2003 Server SP2 & 14\% & 25, 80, 110, 139, \\
 & & & 443, 445, 3389 \\
 \hline
Linux desktop & Ubuntu 8.04 or 8.10 & 27\% & 22 \\
\hline
Linux server & Debian 4.0 & 9\% & 21, 22, 23, 25, \\
 & & & 80, 110, 443 \\
\hline
\end{tabular}
\normalsize
\end{center}
\caption{List of machine types for the test networks}
\label{tab:test_machines}
\end{table}

The network simulator allows us to specify many details about the simulated
machines, so in both networks, the subnetworks attached to the
innermost routers contain four types of machines: Linux desktops and servers,
and Windows desktops and servers. Table \ref{tab:test_machines} shows 
the configuration for each of the four machine types, and the share of each
type in the network. For server cases, each machine randomly removes one
open port from the canonical list shown in the table, so
that all machines are not equal and thus not equally exploitable.

\subsection{Results}

As we expected, both planners generated the same plans in all cases, not
taking into account plans in which goals were composite and the same actions
could be executed in different orders. This is reasonable given
that \Sg uses \Mff as its base planner. We believe that the performance
and scalability results are more interesting, since a valid plan for an
attack path is a satisfactory result in itself.

Figures \ref{fig:scale_time} to \ref{fig:goals_mem} show how running time
and memory consumption scale for both planners, with respect to the four
metrics considered\footnote{Testing was performed on a Core i5 750 2.67
GHz machine with 8 GB of RAM, running 64-bit Ubuntu Linux; the planners
were 32-bit programs.}. Recall that, as explained in Section \ref{sec:pddl},
each exploit maps to many PDDL actions.

\Figure{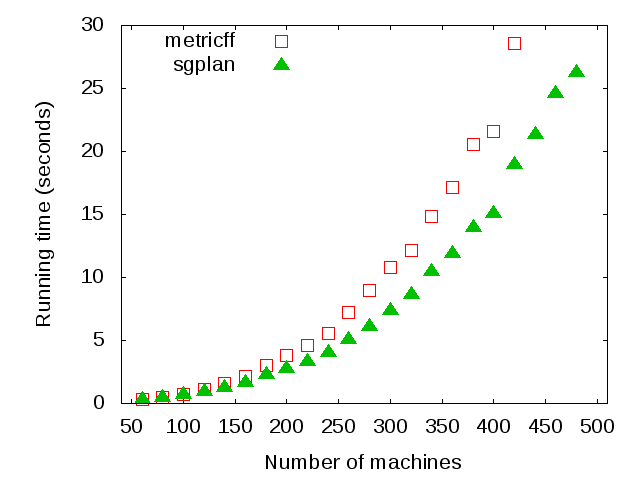}{7cm}{scale_time}
 {Running time, increasing number of machines.
 (Fixed values: 1600 actions, 1 pivoting step).}
\Figure{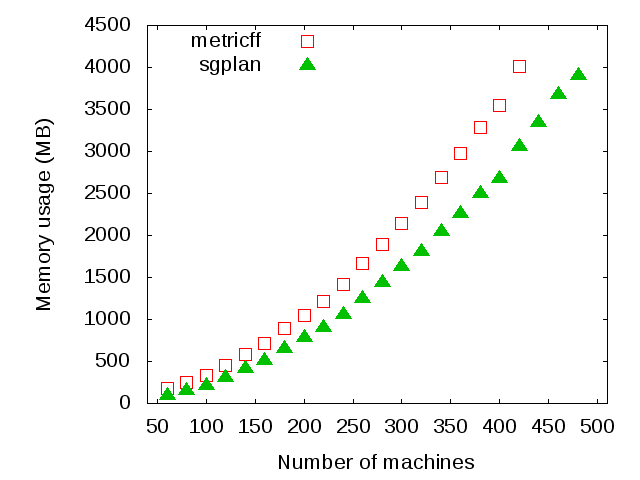}{7cm}{scale_mem}
 {Memory usage, increasing number of machines.}
\Figure{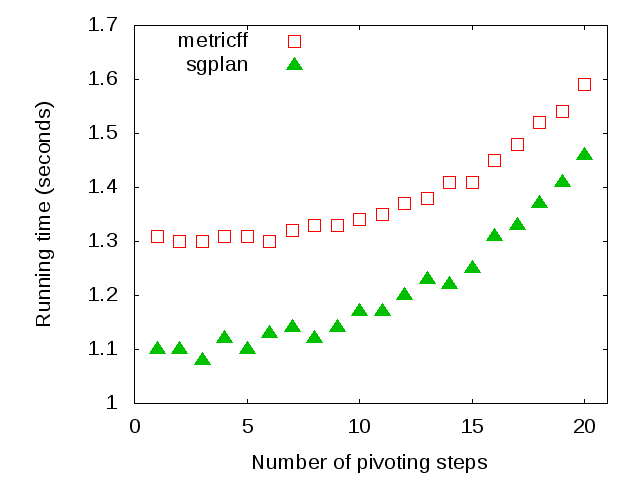}{7cm}{depth_time}
 {Running time, increasing number of pivoting steps.
 (Fixed values: 1600 actions, 120 machines).}
\Figure{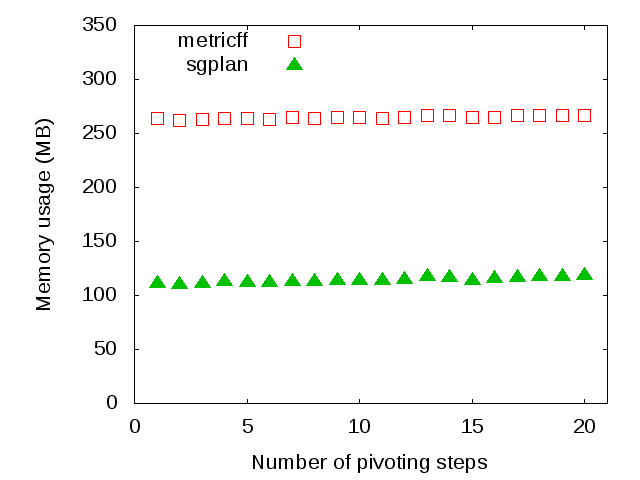}{7cm}{depth_mem}
 {Memory usage, increasing number of pivoting steps.}
\Figure{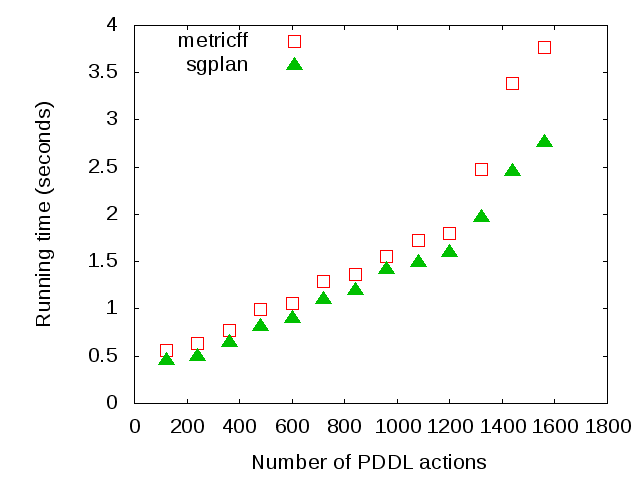}{7cm}{operators_time}
 {Running time, increasing number of actions.
 (Fixed values: 200 machines, 1 pivoting step).}
\Figure{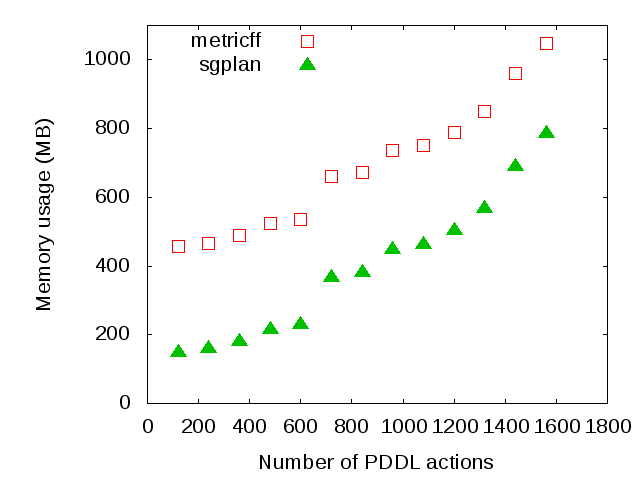}{7cm}{operators_mem}
 {Memory usage, increasing number of actions.}
\Figure{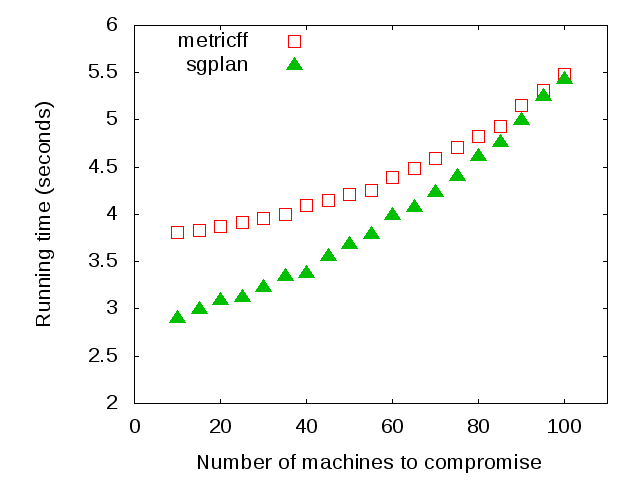}{7cm}{goals_time}
 {Running time, increasing number of predicates in the goal.
 (Fixed values: 200 machines, 1 pivoting step for each compromised machine, 1600 actions).}
\Figure{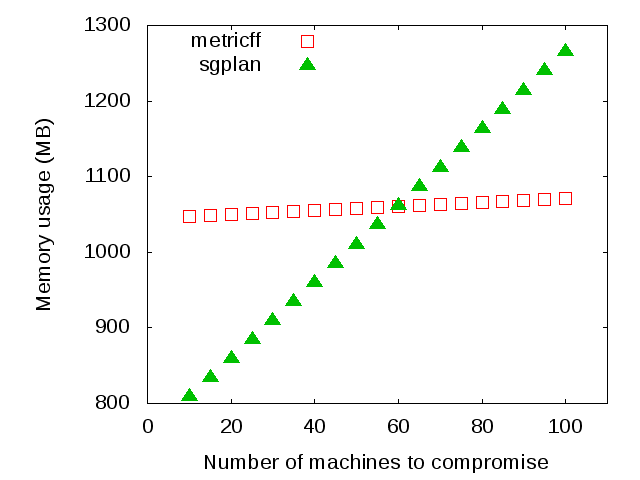}{7cm}{goals_mem}
 {Memory usage, increasing number of predicates in the goal.}

As illustrated by Figures \ref{fig:scale_time} and \ref{fig:scale_mem},
both running time and memory consumption increase superlinearly with the
number of machines in the network. We were not able to find exact specifications
for the time and memory complexities of \Mff or \Sg, though we believe
this is because heuristics make it difficult to calculate a complexity
that holds in normal cases. Nonetheless, our model, coupled with the \Sg
planner, allows to plan an attack in a network with 480 nodes in 25 seconds
and using less than 4 GB of RAM. This makes attack planning practical for
pentests in medium-sized networks.

Moving on to the scalability with regard to the \emph{depth} of the attack
(Figures \ref{fig:depth_time} and \ref{fig:depth_mem}), it was surprising
to verify that memory consumption is constant even as we increase the depth
of the attack to twenty pivoting steps, which generates a plan of more than
sixty steps. Running time increases slowly, although with a small number
of steps the behaviour is less predictable. The model is therefore not
constrained in terms of the number of pivoting steps.

With regard to the number of operators (i.e. exploits) (Figures \ref{fig:operators_time}
and \ref{fig:operators_mem}), both running time and memory consumption
increase almost linearly; however, running time spikes in the largest cases.
Doubling the number of operators, from 720 to 1440 (from 120 to 240 available exploits),
increases running time in 163\% for \Mff and 124\% for \Sg. Memory
consumption, however, increases only 46\% for \Mff, and 87\% for \Sg. In
this context, the number of available exploits is not a limiting factor
for the model.

Interestingly, these three tests also verify many of the claims made by
the authors of \Sg. We see that the constraint partition used by their planner
manages to reduce both running time and memory consumption, in some cases
by significant amounts (like in Figure \ref{fig:depth_mem}).

The results for the individual number of predicates in the overall goal
(Figures \ref{fig:goals_time} and \ref{fig:goals_mem}) are much more surprising. 
While \Sg runs faster than \Mff in most of the cases, \Mff consumes significantly
less memory in almost half of them. We believe that as the goal gets
more complex (the largest case we tested requests the individual compromise
of 100 machines), \Sg's constraint partition strategy turns into a liability,
not allowing a clean separation of the problem into subproblems. By falling
back to \Mff our model can solve, in under 6 seconds and using slightly more
than 1 GB of RAM, attack plans where half of the machines of a 200-machine
network are to be compromised.

\section{Related work}
\label{sec:related}

Work on attack modeling applied to penetration testing had its origin in
the possibility of programmatically controlling pentesting tools such as
Metasploit or Core Impact. This model led to the use of attack graphs.
Earlier work on attack graphs such
as \cite{phillips1998,ritchey2000,sheyner2002} were based on the complete
enumeration of attack states, which grows exponentially with the number
of actions and machines. As we mentioned in Section \ref{sec:introduction}
the survey of \cite{lippmann2005} shows that the major limitations of past
studies of attack graphs is their lack of scalability to medium-sized networks.

One notable exception is the Topological Vulnerability Analysis (TVA) project conducted in 
George Mason University described in \cite{jajodia2005,noel2005understanding,noel2009}
and other papers, which has been designed to work in real-size networks.
The main differences between our approach and TVA are the following:

\begin{itemize}

\item{ {\bf Input.} In TVA the model is populated with information from
third party vulnerability scanners such as Nessus, Retina and FoundScan,
from databases of vulnerabilities such as CVE and OSVDB and other software.
All this information has to be integrated, and will suffer from the drawbacks
of each information source, in particular from the false positives generated
by the vulnerability scanners about potential vulnerabilities.

In our approach the conceptual model and the information about the target
network are extracted from a consistent source: the pentesting framework exploits
and workspace. The vulnerability information of an exploit is very precise:
the attacker can execute it in the real network to actually compromise systems.
}

\item{ {\bf Monotonicity.} TVA assumes that the attacker's control over
the network is monotonic \cite{ammann2002}. In particular, this implies
that TVA cannot model Denial-of-Service (DoS) attacks, or the fact that
an unsuccessful exploit may crash the target service or machine. It is
interesting to remark that the monotonicity assumption is the same used
by FF \cite{hoffmann2001ff} to create a relaxed version of the planning
problem, and use it as a heuristic
to guide the search through the attack graph. By relying on the planner
to do the search, we do not need to make this restrictive assumption.
}
\end{itemize}

\section{Summary and Future Work}
\label{sec:conclusion}

\cite{building2003} proposed a model of computer network attacks which
was designed to be realistic from an attacker's point of view.
We have shown in this paper that this model scales up to medium-sized networks:
it can be used to automate attacks (and penetration tests) against networks
with hundreds of machines.

The solution presented shows that it is not necessary to build the complete
attack graph (one of the major limitations of earlier attack graph studies).
Instead we rely on planners such as \Mff and \Sg to selectively explore
the state space in order to find attack paths.

We have successfully integrated these planners with a pentesting framework,
which allowed us to execute and validate the resulting plans against a
test bench of scenarios. We presented the details of how to transform the
information contained in the pentesting tool to the planning domain\footnote{Our
implementation uses Core Impact, but the same ideas can be extended to
other tools such as the open-source project Metasploit.}.

One important question that remains as future work on this subject is how
to deal with incomplete knowledge of the target network. The architecture that
we presented supports running non-classical planners, so one possible approach
is to use probabilistic planning techniques, where actions have different
outcomes with associated probabilities. For example, a step of the attack
plan could be to discover the operating system details of a particular host,
so the outcome of this action would be modeled as a discrete probability
distribution.

Another approach would be to build a ``metaplanner'' that generates hypotheses
with respect to the missing bits of information about the network, and
uses the planner to test those hypotheses. Continuing the previous example,
the metaplanner would assume that the operating system of the host was
Windows and request the planner to compromise it as such. The metaplanner
would then test the resulting plan in the real network, and verify or discard
the hypothesis.

%

\end{document}

%% file: plan.tex

\scriptsize 
\begin{verbatim}
 0: Mark_as_compromised localagent localhost
 1: IP_connect localhost 10.0.1.1
 2: TCP_connect localhost 10.0.1.1 port80
 3: Phpmyadmin Server_databases Remote Code Execution
        localhost 10.0.1.1
 4: Mark_as_compromised 10.0.1.1 high_privileges
 ...
14: Mark_as_compromised 10.0.4.2 high_privileges
15: IP_connect 10.0.4.2 10.0.5.12
16: TCP_connect 10.0.4.2 10.0.5.12 port445
17: Novell Client NetIdentity Agent Buffer Overflow
        10.0.4.2 10.0.5.12
18: Mark_as_compromised 10.0.5.12 high_privileges
\end{verbatim}
\normalsize